\documentclass[%
reprint,
superscriptaddress,
showpacs,
 amsmath,amssymb,
 aps,
]{revtex4-1}

\usepackage{graphicx}
\usepackage{dcolumn}
\usepackage{bm}
\usepackage{amsmath}
\usepackage{autobreak}
\usepackage{hyperref}
\usepackage[mathlines]{lineno}
\usepackage{natbib}
\usepackage{multirow}
\usepackage{mathtools}
\usepackage{color}

\usepackage{booktabs}

\def\bb{0\nu\beta\beta}

\def\hfconstrain#1#2{T_{1/2}^{0\nu}\ > \ {#1}\times10^{#2} \ \mathrm{yr}}
\def\Thf{T_{1/2}^{0\nu}}

\begin{document}

\preprint{APS/123-QED}

\title{Searching for $^{76}$Ge neutrinoless double beta decay with the CDEX-1B experiment}

\author{B.~T.~Zhang}
\affiliation{Key Laboratory of Particle and Radiation Imaging (Ministry of Education) and Department of Engineering Physics, Tsinghua University, Beijing 100084}
\author{J.~Z.~Wang}
\affiliation{Key Laboratory of Particle and Radiation Imaging (Ministry of Education) and Department of Engineering Physics, Tsinghua University, Beijing 100084}
\author{L.~T.~Yang}\altaffiliation [Corresponding author: ]{yanglt@mail.tsinghua.edu.cn}
\affiliation{Key Laboratory of Particle and Radiation Imaging (Ministry of Education) and Department of Engineering Physics, Tsinghua University, Beijing 100084}
\author{Q. Yue}\altaffiliation [Corresponding author: ]{yueq@mail.tsinghua.edu.cn}
\affiliation{Key Laboratory of Particle and Radiation Imaging (Ministry of Education) and Department of Engineering Physics, Tsinghua University, Beijing 100084}

\author{K.~J.~Kang}
\affiliation{Key Laboratory of Particle and Radiation Imaging (Ministry of Education) and Department of Engineering Physics, Tsinghua University, Beijing 100084}
\author{Y.~J.~Li}
\affiliation{Key Laboratory of Particle and Radiation Imaging (Ministry of Education) and Department of Engineering Physics, Tsinghua University, Beijing 100084}

\author{H.~P.~An}
\affiliation{Key Laboratory of Particle and Radiation Imaging (Ministry of Education) and Department of Engineering Physics, Tsinghua University, Beijing 100084}
\affiliation{Department of Physics, Tsinghua University, Beijing 100084}

\author{Greeshma~C.}
\altaffiliation{Participating as a member of TEXONO Collaboration}
\affiliation{Institute of Physics, Academia Sinica, Taipei 11529}

\author{J.~P.~Chang}
\affiliation{NUCTECH Company, Beijing 100084}

\author{Y.~H.~Chen}
\affiliation{YaLong River Hydropower Development Company, Chengdu 610051}
\author{J.~P.~Cheng}
\affiliation{Key Laboratory of Particle and Radiation Imaging (Ministry of Education) and Department of Engineering Physics, Tsinghua University, Beijing 100084}
\affiliation{College of Nuclear Science and Technology, Beijing Normal University, Beijing 100875}
\author{W.~H.~Dai}
\affiliation{Key Laboratory of Particle and Radiation Imaging (Ministry of Education) and Department of Engineering Physics, Tsinghua University, Beijing 100084}
\author{Z.~Deng}
\affiliation{Key Laboratory of Particle and Radiation Imaging (Ministry of Education) and Department of Engineering Physics, Tsinghua University, Beijing 100084}
\author{C.~H.~Fang}
\affiliation{College of Physics, Sichuan University, Chengdu 610065}
\author{X.~P.~Geng}
\affiliation{Key Laboratory of Particle and Radiation Imaging (Ministry of Education) and Department of Engineering Physics, Tsinghua University, Beijing 100084}
\author{H.~Gong}
\affiliation{Key Laboratory of Particle and Radiation Imaging (Ministry of Education) and Department of Engineering Physics, Tsinghua University, Beijing 100084}
\author{Q.~J.~Guo}
\affiliation{School of Physics, Peking University, Beijing 100871}
\author{X.~Y.~Guo}
\affiliation{YaLong River Hydropower Development Company, Chengdu 610051}
\author{L.~He}
\affiliation{NUCTECH Company, Beijing 100084}
\author{S.~M.~He}
\affiliation{YaLong River Hydropower Development Company, Chengdu 610051}
\author{J.~W.~Hu}
\affiliation{Key Laboratory of Particle and Radiation Imaging (Ministry of Education) and Department of Engineering Physics, Tsinghua University, Beijing 100084}
\author{H.~X.~Huang}
\affiliation{Department of Nuclear Physics, China Institute of Atomic Energy, Beijing 102413}
\author{T.~C.~Huang}
\affiliation{Sino-French Institute of Nuclear and Technology, Sun Yat-sen University, Zhuhai 519082}
\author{H.~T.~Jia}
\affiliation{College of Physics, Sichuan University, Chengdu 610065}
\author{X.~Jiang}
\affiliation{College of Physics, Sichuan University, Chengdu 610065}

\author{S.~Karmakar}
\altaffiliation{Participating as a member of TEXONO Collaboration}
\affiliation{Institute of Physics, Academia Sinica, Taipei 11529}

\author{H.~B.~Li}
\altaffiliation{Participating as a member of TEXONO Collaboration}
\affiliation{Institute of Physics, Academia Sinica, Taipei 11529}
\author{J.~M.~Li}
\affiliation{Key Laboratory of Particle and Radiation Imaging (Ministry of Education) and Department of Engineering Physics, Tsinghua University, Beijing 100084}
\author{J.~Li}
\affiliation{Key Laboratory of Particle and Radiation Imaging (Ministry of Education) and Department of Engineering Physics, Tsinghua University, Beijing 100084}
\author{Q.~Y.~Li}
\affiliation{College of Physics, Sichuan University, Chengdu 610065}
\author{R.~M.~J.~Li}
\affiliation{College of Physics, Sichuan University, Chengdu 610065}
\author{X.~Q.~Li}
\affiliation{School of Physics, Nankai University, Tianjin 300071}
\author{Y.~L.~Li}
\affiliation{Key Laboratory of Particle and Radiation Imaging (Ministry of Education) and Department of Engineering Physics, Tsinghua University, Beijing 100084}
\author{Y.~F.~Liang}
\affiliation{Key Laboratory of Particle and Radiation Imaging (Ministry of Education) and Department of Engineering Physics, Tsinghua University, Beijing 100084}
\author{B.~Liao}
\affiliation{College of Nuclear Science and Technology, Beijing Normal University, Beijing 100875}
\author{F.~K.~Lin}
\altaffiliation{Participating as a member of TEXONO Collaboration}
\affiliation{Institute of Physics, Academia Sinica, Taipei 11529}
\author{S.~T.~Lin}
\affiliation{College of Physics, Sichuan University, Chengdu 610065}
\author{J.~X.~Liu}
\affiliation{Key Laboratory of Particle and Radiation Imaging (Ministry of Education) and Department of Engineering Physics, Tsinghua University, Beijing 100084}
\author{S.~K.~Liu}
\affiliation{College of Physics, Sichuan University, Chengdu 610065}
\author{Y.~D.~Liu}
\affiliation{College of Nuclear Science and Technology, Beijing Normal University, Beijing 100875}
\author{Y.~Liu}
\affiliation{College of Physics, Sichuan University, Chengdu 610065}
\author{Y.~Y.~Liu}
\affiliation{College of Nuclear Science and Technology, Beijing Normal University, Beijing 100875}
\author{Z.~Z.~Liu}
\affiliation{Key Laboratory of Particle and Radiation Imaging (Ministry of Education) and Department of Engineering Physics, Tsinghua University, Beijing 100084}
\author{H.~Ma}
\affiliation{Key Laboratory of Particle and Radiation Imaging (Ministry of Education) and Department of Engineering Physics, Tsinghua University, Beijing 100084}
\author{Y.~C.~Mao}
\affiliation{School of Physics, Peking University, Beijing 100871}
\author{Q.~Y.~Nie}
\affiliation{Key Laboratory of Particle and Radiation Imaging (Ministry of Education) and Department of Engineering Physics, Tsinghua University, Beijing 100084}
\author{J.~H.~Ning}
\affiliation{YaLong River Hydropower Development Company, Chengdu 610051}
\author{H.~Pan}
\affiliation{NUCTECH Company, Beijing 100084}
\author{N.~C.~Qi}
\affiliation{YaLong River Hydropower Development Company, Chengdu 610051}
\author{J.~Ren}
\affiliation{Department of Nuclear Physics, China Institute of Atomic Energy, Beijing 102413}
\author{X.~C.~Ruan}
\affiliation{Department of Nuclear Physics, China Institute of Atomic Energy, Beijing 102413}

\author{Z.~She}
\affiliation{Key Laboratory of Particle and Radiation Imaging (Ministry of Education) and Department of Engineering Physics, Tsinghua University, Beijing 100084}
\author{M.~K.~Singh}
\altaffiliation{Participating as a member of TEXONO Collaboration}
\affiliation{Institute of Physics, Academia Sinica, Taipei 11529}
\affiliation{Department of Physics, Banaras Hindu University, Varanasi 221005}
\author{T.~X.~Sun}
\affiliation{College of Nuclear Science and Technology, Beijing Normal University, Beijing 100875}
\author{C.~J.~Tang}
\affiliation{College of Physics, Sichuan University, Chengdu 610065}
\author{W.~Y.~Tang}
\affiliation{Key Laboratory of Particle and Radiation Imaging (Ministry of Education) and Department of Engineering Physics, Tsinghua University, Beijing 100084}
\author{Y.~Tian}
\affiliation{Key Laboratory of Particle and Radiation Imaging (Ministry of Education) and Department of Engineering Physics, Tsinghua University, Beijing 100084}
\author{G.~F.~Wang}
\affiliation{College of Nuclear Science and Technology, Beijing Normal University, Beijing 100875}
\author{L.~Wang}
\affiliation{Department of  Physics, Beijing Normal University, Beijing 100875}
\author{Q.~Wang}
\affiliation{Key Laboratory of Particle and Radiation Imaging (Ministry of Education) and Department of Engineering Physics, Tsinghua University, Beijing 100084}
\affiliation{Department of Physics, Tsinghua University, Beijing 100084}
\author{Y.~F.~Wang}
\affiliation{Key Laboratory of Particle and Radiation Imaging (Ministry of Education) and Department of Engineering Physics, Tsinghua University, Beijing 100084}
\author{Y.~X.~Wang}
\affiliation{School of Physics, Peking University, Beijing 100871}
\author{H.~T.~Wong}
\altaffiliation{Participating as a member of TEXONO Collaboration}
\affiliation{Institute of Physics, Academia Sinica, Taipei 11529}
\author{S.~Y.~Wu}
\affiliation{YaLong River Hydropower Development Company, Chengdu 610051}
\author{Y.~C.~Wu}
\affiliation{Key Laboratory of Particle and Radiation Imaging (Ministry of Education) and Department of Engineering Physics, Tsinghua University, Beijing 100084}
\author{H.~Y.~Xing}
\affiliation{College of Physics, Sichuan University, Chengdu 610065}
\author{R. Xu}
\affiliation{Key Laboratory of Particle and Radiation Imaging (Ministry of Education) and Department of Engineering Physics, Tsinghua University, Beijing 100084}
\author{Y.~Xu}
\affiliation{School of Physics, Nankai University, Tianjin 300071}
\author{T.~Xue}
\affiliation{Key Laboratory of Particle and Radiation Imaging (Ministry of Education) and Department of Engineering Physics, Tsinghua University, Beijing 100084}
\author{Y.~L.~Yan}
\affiliation{College of Physics, Sichuan University, Chengdu 610065}

\author{N.~Yi}
\affiliation{Key Laboratory of Particle and Radiation Imaging (Ministry of Education) and Department of Engineering Physics, Tsinghua University, Beijing 100084}
\author{C.~X.~Yu}
\affiliation{School of Physics, Nankai University, Tianjin 300071}
\author{H.~J.~Yu}
\affiliation{NUCTECH Company, Beijing 100084}
\author{J.~F.~Yue}
\affiliation{YaLong River Hydropower Development Company, Chengdu 610051}
\author{M.~Zeng}
\affiliation{Key Laboratory of Particle and Radiation Imaging (Ministry of Education) and Department of Engineering Physics, Tsinghua University, Beijing 100084}
\author{Z.~Zeng}
\affiliation{Key Laboratory of Particle and Radiation Imaging (Ministry of Education) and Department of Engineering Physics, Tsinghua University, Beijing 100084}
\author{F.~S.~Zhang}
\affiliation{College of Nuclear Science and Technology, Beijing Normal University, Beijing 100875}
\author{L.~Zhang}
\affiliation{College of Physics, Sichuan University, Chengdu 610065}
\author{Z.~H.~Zhang}
\affiliation{Key Laboratory of Particle and Radiation Imaging (Ministry of Education) and Department of Engineering Physics, Tsinghua University, Beijing 100084}
\author{Z.~Y.~Zhang}
\affiliation{Key Laboratory of Particle and Radiation Imaging (Ministry of Education) and Department of Engineering Physics, Tsinghua University, Beijing 100084}
\author{K.~K.~Zhao}
\affiliation{College of Physics, Sichuan University, Chengdu 610065}
\author{M.~G.~Zhao}
\affiliation{School of Physics, Nankai University, Tianjin 300071}
\author{J.~F.~Zhou}
\affiliation{YaLong River Hydropower Development Company, Chengdu 610051}
\author{Z.~Y.~Zhou}
\affiliation{Department of Nuclear Physics, China Institute of Atomic Energy, Beijing 102413}
\author{J.~J.~Zhu}
\affiliation{College of Physics, Sichuan University, Chengdu 610065}

\collaboration{CDEX Collaboration}
\noaffiliation

\date{\today}

\begin{abstract}  We operated a p-type point contact high purity germanium (PPCGe) detector (CDEX-1B, 1.008 kg) in the China Jinping Underground Laboratory (CJPL) for 500.3 days to search for neutrinoless double beta ($\bb$) decay of $^{76}$Ge. A total of 504.3 kg$\cdot$day effective exposure data was accumulated. The anti-coincidence and the multi/single-site event (MSE/SSE) discrimination methods were used to suppress the background in the energy region of interest (ROI, 1989--2089 keV for this work) with a factor of 23. A background level of 0.33 counts/(keV$\cdot$kg$\cdot$yr) was realized. The lower limit on the half life of $^{76}$Ge $\bb$ decay was constrained as $T_{1/2}^{0\nu}\ > \ {1.0}\times 10^{23}\ \rm yr\ (90\% \ C.L.)$, corresponding to the upper limits on the effective Majorana neutrino mass: $\langle m_{\beta\beta}\rangle < $ 3.2--7.5$\ \mathrm{eV}$. 
\end{abstract}

\maketitle

\section{Introduction} 
The matter-antimatter asymmetry of the Universe is one of the greatest mysteries in cosmology and particle physics~\cite{leptogenesis,GERDA2020}. Leptogenesis is a leading theory that explains this asymmetry through the violation of lepton number conservation~\cite{leptogenesis,universe7090341,RevModPhys.95.025002}, implying that neutrinos include a Majorana mass component and act as their own antiparticles. In principle, the Majorana nature of neutrinos and lepton number violation can be tested by observing a hypothetical nuclear process, termed as neutrinoless double-beta ($\bb$) decay~\cite{zel1981study}. The search for $\bb$ decay has been deemed as the most promising approach to probe the Majorana nature of neutrinos, and its observation can provide direct evidence for a process beyond the Standard Model, which violates lepton number conservation and constrains the absolute mass scale of neutrinos~\cite{leptogenesis,status,HP2015,EXO2019,GERDA2020,MJD2023}.

Currently, researchers worldwide are investing efforts to search for this rare decay in various isotopes, such as $^{76}$Ge (CDEX~\cite{CDEX2017,DWH2022}, GERDA~\cite{GERDA2020} and M\textsc{ajorana} D\textsc{emonstrator}~\cite{MJD2023}), $^{130}$Te (CUORE~\cite{CUORE2020}, SNO+~\cite{SNO+}), $^{136}$Xe (KamLAND-Zen~\cite{KamLAND-Zen2023}, EXO~\cite{EXO2019}), and $^{100}$Mo (NEMO-3~\cite{NEMO2015}). In $\bb$ decay process, two neutrons in a nucleus are converted into two protons with the emission of two electrons. As the recoil of the nucleus is negligible, the two electrons carry all the decay energy. The key experimental signature of the $\bb$ decay corresponds to a peak centered at the Q value ($Q_{\beta\beta}$) of the decay. The high purity germanium (HPGe) detector, serving as both target nuclei and detector, is an ideal medium for detecting $\bb$ decays of $^{76}$Ge because of its high energy resolution, low internal background, and high detection efficiency~\cite{soma2016,Mjd2018,DWH2022,BEGeGERDA,ICPCGERDA,ICPCLEGEND}. The $^{76}$Ge-based $\bb$ experiments have been conducted for many years, and the experimental sensitivities have been continuously improved. To date, the best half-life limit for $^{76}$Ge is afforded by GERDA: $\hfconstrain{1.8}{26}$; the corresponding upper limit for effective Majorana neutrino mass ($ m_{\beta\beta}$) is in the range of 79--180 meV~\cite{GERDA2020} (With the latest values of nuclear matrix element, 2.66--6.34~\cite{RevModPhys.95.025002}, this range is 75--180 meV).

Based on the China Jinping Underground Laboratory (CJPL)~\cite{cjpl}, CDEX collaboration is committed to employing HPGe detectors for dark matter direct detection and $^{76}$Ge $\bb$ decay searches. The CDEX has completed two stages of dedicated dark matter search experiments, CDEX-1 and CDEX-10 with p-type point contact (PPC) HPGe detectors~\cite{CDEX2017,CDEX2014,CDEX2016,CDEX2018,cdex102018,CDEX2019,CDEX2019Liu,CDEX2020,XuR2022,ZhangZY2022}. In this study, we explored the use of a PPCGe in the CDEX-1B experiment for $^{76}$Ge $\bb$ detection as a feasibility study for its use in the future CDEX-300$\nu$ experiment. By tuning the data acquisition system and applying two active background suppression technologies (pulse shape discrimination and anti-coincidence veto), we ran the CDEX-1B PPCGe detector in $\bb$ search mode for over 500 days and presented a new and best $^{76}$Ge $\bb$ result within the CDEX experiments. This expands the physical objectives of CDEX experiments and explores the feasibility of using large high-purity germanium arrays to simultaneously conduct dark matter and $^{76}$Ge $\bb$ experiments in the future.

\section{Experimental setup}
The CDEX-1B detector is a p-type point contact high purity germanium (PPCGe) detector operated in CJPL with a rock depth of 2400 m. Owing to the ultralow cosmic ray flux in CJPL, the background radiation due to cosmic rays is reduced to a negligible level~\cite{CDEX2013}. The CDEX-1B PPCGe detector is a natural germanium detector with a $^{76}$Ge abundance of 7.83\% and total mass of 1.008 kg. Ge crystal is a cylinder with a height of 62.3 mm and diameter of 62.1 mm, whose signals are read out from the P+ point contact at the bottom. The dead-layer thickness was evaluated as  $0.88\pm{0.12} \ \mathrm{mm}$~\cite{CDEX2018,MJL2017}. The energy resolution of a single site event (SSE) at $Q_{\beta\beta}=2039 \ \mathrm{keV}$ is 2.76 keV (full width at half maximum, FHWM). To operate the detector in an extremely low background environment, we constructed a complex shielding system, including passive shielding (from inside out: with 20 cm of copper, 20 cm of borated polyethylene, 20 cm of lead, and 1 m of polyethylene) as well as active shielding (NaI anti-coincidence detector) to reduce the external background~\cite{CDEX2019,CDEX2018}. The output signal from the P+ point contact electrode of germanium crystal was fed into a pulsed reset preamplifier (Canberra PSC-954P), which is fanned out to provide four outputs. Two of them were distributed into Canberra 2026 shaping amplifiers. The other two outputs were loaded to Canberra 2111 timing amplifiers, which maintained relatively accurate information of the rising edge of pulses. All output signals were digitalized using a 100-MHz flash analog-to-digital converter (FADC; CAEN V1724) with 14-bit accuracy. The waveform sampling window was 120 $\mathrm{\mu s}$~\cite{CDEX2018}.

The CDEX-1B detector was used to detect dark matter by operating for three periods (runs I, II, and III), and it achieved a series of physics results~\cite{CDEX2019Liu,CDEX2018,CDEX2019}. To conduct the $\bb$ experiment, we adjusted the dynamic range of the preamplifier to cover high energy region for the $\bb$ signal detection. In this study, we used datasets from two runs, i.e., IV and V, which were acquired between February 6, 2021 and April 28, 2021, and August 9, 2021 to January 8, 2023, respectively. After excluding the system maintenance period, the total effective runtime was 500.3 days. Thus, the effective exposure, defined as the product of the running time and detector mass (here, 1.008 kg), was 504.3 kg$\cdot$day. 

\begin{figure}[!htbp]
\centering
\includegraphics[width=\linewidth]{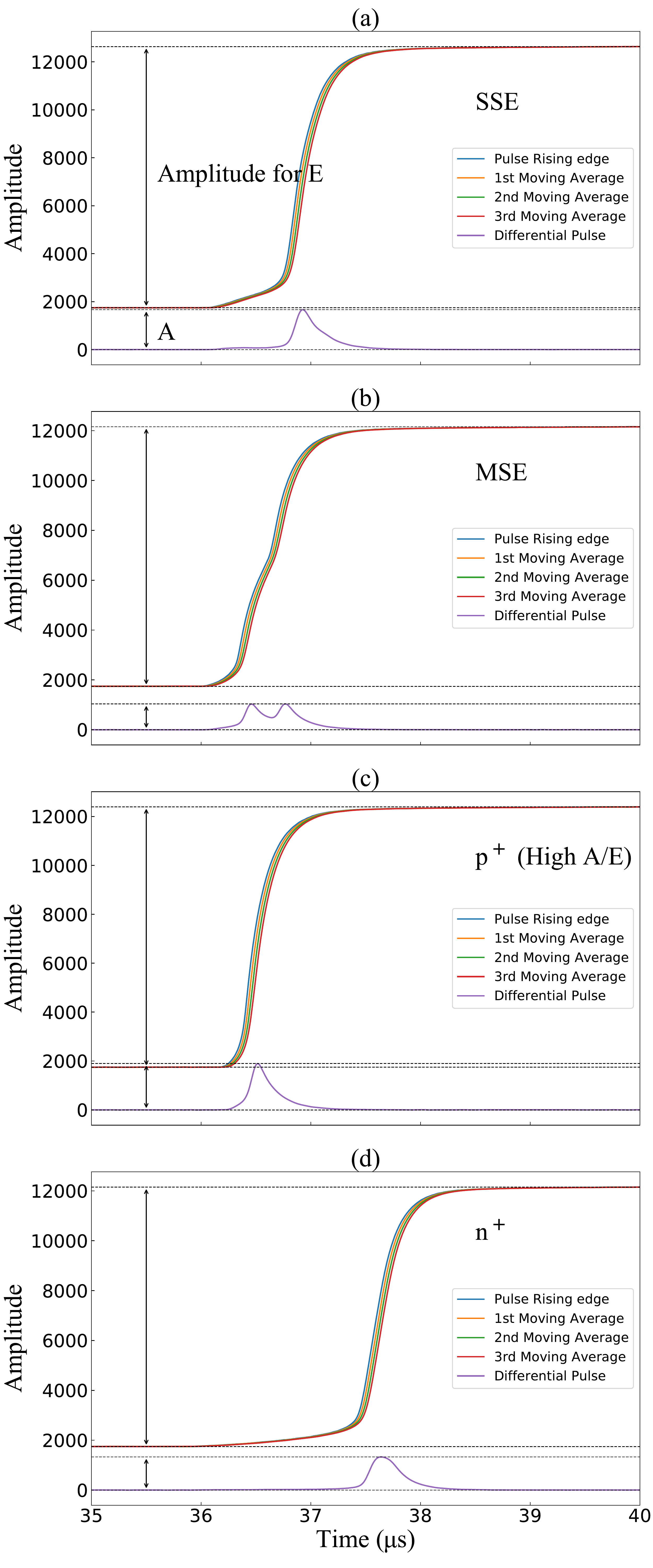}
\caption{Pulse information extraction process; (a) single-site event, (b) multisite event, (c) high A/E event, and (d) n+ event. The energies of the events in the four figures are 2102.6, 2008.4, 2054.3, and 2012.5 keV, respectively.}
\label{fig:Pulse_info}
\end{figure}

\section{Data analysis}
The extraction of pulse shape information involves three steps: (1) for denoising, we performed a triple moving average with a 0.05-$\mu$s window (5 sampling points for the 100-MHz sampling rate) on the fast amplifier pulse; (2) the maximum amplitude of the denoised pulse was used for energy calibration; and (3) the current pulse was extracted from the denoised pulse by using a differential filter, and the maximum amplitude of the current pulse (A) was used in the subsequent A/E method.

The pulse information extraction process is illustrated in Fig.~\ref{fig:Pulse_info}. The moving-average filtering with a window width of 50 ns was performed three times on the original pulse according to Eq.~\ref{eq:MOVAVG}, where $P_k^{(n)}$ denotes the amplitude of the $k$-th sampling point of the pulse after $n$ iterations and $L=5$ (10 ns for each point) denotes the width of the moving-average window. The maximum amplitude was utilized for energy calibration.
\begin{equation}
P_i^{(n)} = \begin{cases}
\frac{1}{L} \sum_{k=i-L+1}^{k=i} P_k^{(n-1)} &  i \geq {L-1}, \\ 
\frac{1}{i+1} \sum_{k=0}^{k=i} P_k^{(n-1)} & 
i < {L-1}\end{cases}
\label{eq:MOVAVG}
\end{equation}

The current pulse was extracted from the denoised pulse by using a differential filter (Eq.~\ref{eq:partial}), where $L=5$ denotes the differential step size. Parameter A/E was defined as the ratio of the maximum current amplitude (A) to the reconstructed energy (E).

\begin{equation}
I_i = \begin{cases}
P_i^{(3)}-P_{i-L}^{(3)} &  i \geq {L}, \\ 
0 & i < {L}\end{cases}
\label{eq:partial}
\end{equation}

\begin{figure}[!tbp]
\centering
\includegraphics[width=\linewidth]{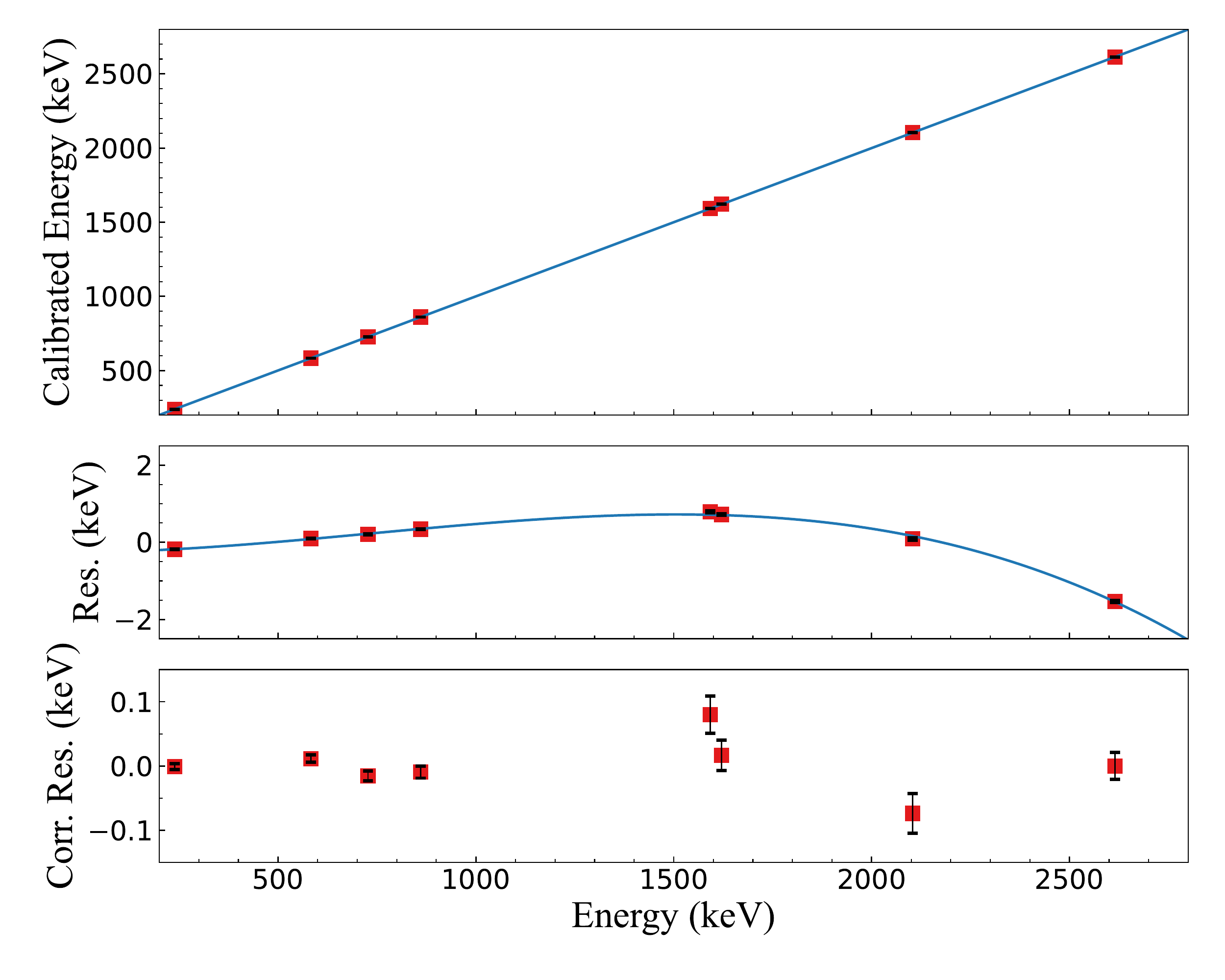}
\caption{
(Top) Linear relationship between the calibrated energy and real energy. (Middle) Residual of the calibrated energy with respect to real energy. The residual was fitted using a cubic polynomial, and the fitting results are used to correct the linear calibrated results. (Bottom) Corrected residual; the horizontal axis represents real energy. After correction, the corrected residual was less than 0.1 keV.}
\label{fig:Th228_Cali}
\end{figure}

Between runs IV and V of data acquisition, we disassembled the external shields of the CDEX-1B detector, removed the NaI detector, and conducted a calibration experiment with a $^{228}$Th radiation source. Figure~\ref{fig:Th228_Cali} shows the energy calibration result. We selected eight gamma rays of 238.63 keV ($^{208}$Tl), 583.19 keV ($^{208}$Tl), 727.33 keV ($^{212}$Bi), 860.56 keV ($^{208}$Tl), 1592.50 keV ($^{208}$Tl), 1620.50 keV ($^{212}$Bi), 2104.50 keV ($^{208}$Tl single-escape peak), and 2614.51 keV ($^{208}$Tl) energies for energy calibration and used a cubic polynomial function to reduce the residuals of the calibrated energy~\cite{GERDACalibration}. After correction, the maximum residual was less than 0.1 keV. After the calibration, the external shield and NaI detector were restored to their original states. 

We should emphasize that the calibrated energy in this study was only used to evaluate energy resolution and A/E cut parameters. The same calibration method was applied to each exposure dataset separately to guarantee the stability of calibration during the long-term operation, with five energy peaks in the low background spectrum: 238.63 keV ($^{208}$Tl), 583.19 keV ($^{208}$Tl), 1460.75 keV ($^{40}$K), 1764.49 keV ($^{214}$Bi), and 2614.51 keV ($^{208}$Tl). The corrected residual was maintained below 0.2 keV.

\begin{figure}[!tbp]
\includegraphics[width=\linewidth]{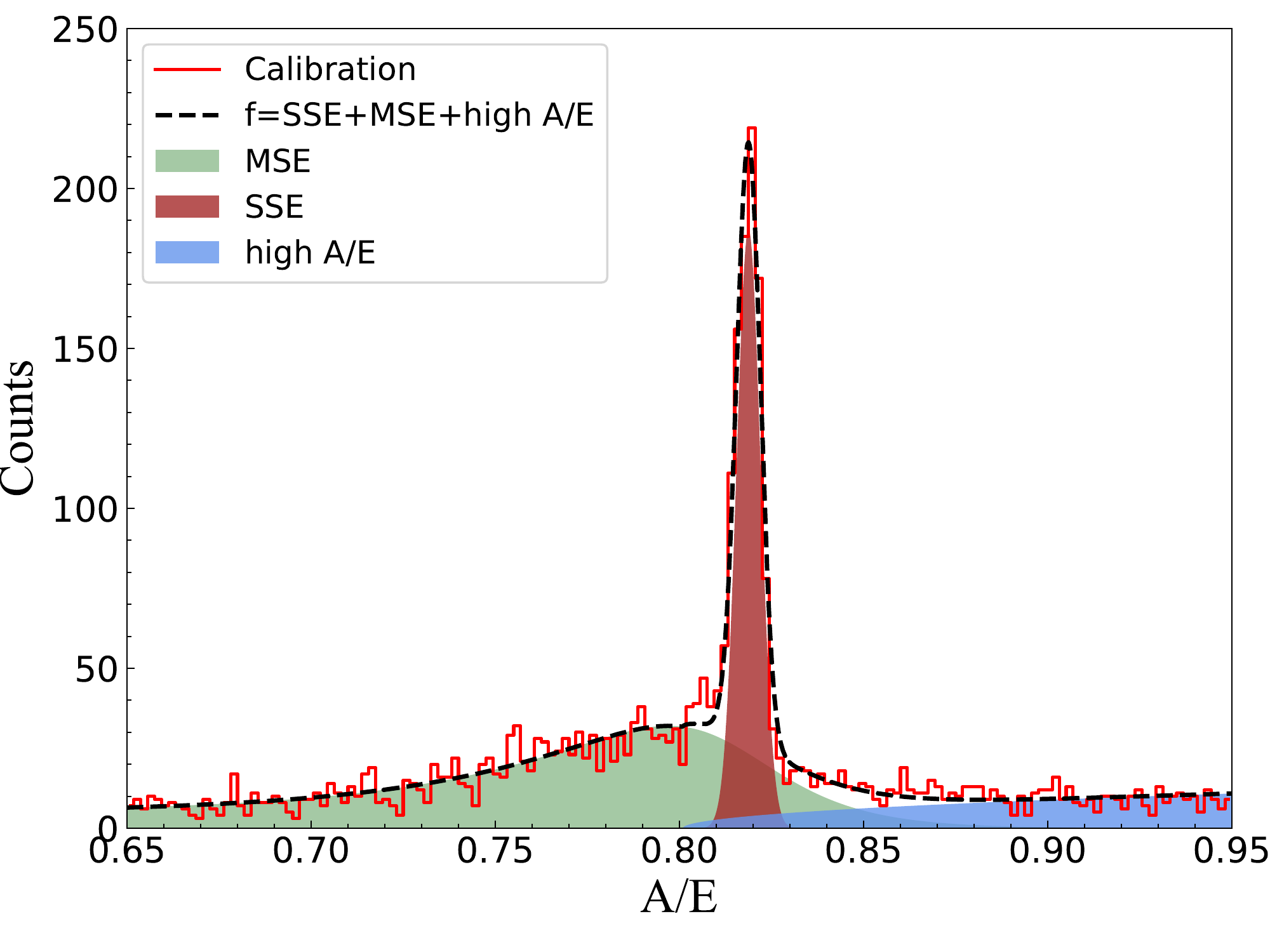}
\caption{
A/E distribution of $^{228}$Th calibration events in energy intervals of [1745 keV, 1775 keV], including three major components: MSEs, SSEs, and high A/E events. The A/E values of multiple components follow different distribution forms, and the A/E of the SSEs follows the Gaussian distribution. Dashed lines denote the fitting result of the A/E distribution using Eq.~\ref{equa1}; green, red, and blue shadows represent the MSE, SSE, and high A/E event components, respectively.
}
\label{fig:Cali_AE_C1B_sketch}
\end{figure}

The $\bb$ process only emits two electrons, and the range of these two electrons in a germanium crystal is approximately 1 mm. Therefore, most of the energy of a $\bb$ event will be deposited in a point-like region, and $\bb$ event is regarded as an SSE. The majority of the background events are multisite events (MSE), such as the multiple Compton scattering events. Thus, distinguishing SSEs from MSEs and discarding MSEs as background events, the background level in the signal region can be effectively reduced. For the MSE, as the energy is deposited in multiple spatial points, the multiple charge carrier clusters will reach the P+ point contact at different times. However, an SSE with equivalent energy comprises only one carrier cluster, and the maximal current amplitude, generated when the hole carrier cluster reaches the P+ point contact, is higher than that of the MSE. Therefore, we can identify the nature of the event, i.e., SSE or MSE, based on the ratio of the maximal current value A to the energy E of the event (A/E)~\cite{GERDAPSD,GERDAPSD2022,MjdPSD2019}. Additionally, certain events exhibited higher A/E values with respect to typical SSEs, which are referred to as  high A/E events. These types of events can be segmented into two types according to their origin. First, the events occur proximate to the point electrode, during which the hole carriers will reach the P-point electrode in an extremely short period and generate a current peak. At this instant, the electron carriers are still drifting, and the current signal generated by the electrons will be superimposed on the current peak, thereby increasing the maximal current value. Then, the event occurs proximate to the passivation layer. In this event, the trajectory of charge carrier holes passes through the surface region near the passivation layer, where the gradient of the weight potential field is larger, thereby leading to a higher A/E value in the induced signal pulse~\cite{GERDAPSD_2009,GERDAPSD2022,MJDPSD2011}. These two types of events constitute high A/E events. Figure~\ref{fig:Cali_AE_C1B_sketch} shows the A/E distribution of $^{228}$Th calibration events in an energy interval of [1745 keV, 1775 keV], and the A/E distributions of the three abovementioned types of events are illustrated in different colors.

\begin{figure}[!tbp]
\includegraphics[width=\linewidth]{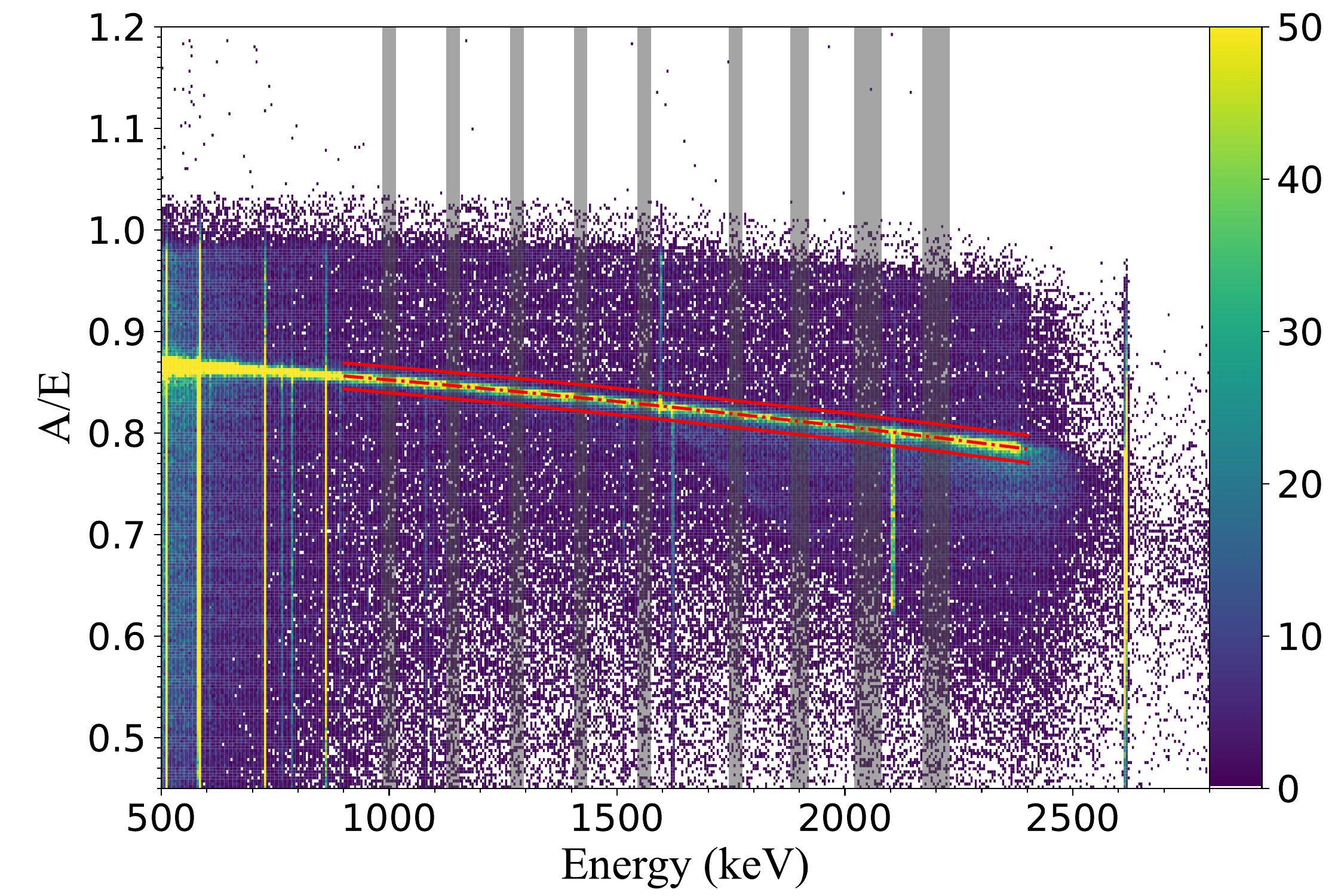}
\caption{
Two-dimensional distribution diagram of A/E vs. the energy of the $^{228}$Th calibration data. The upper and lower red dashed lines correspond to $\mu(E)+4\sigma(E)$ and $\mu(E)-4\sigma(E)$ threshold, respectively. The SSEs exhibit a band distribution that decreases gradually with the energies. The major component of the double escape peak (DEP; 1592.50 keV) of 2614.51 keV $\gamma$-rays ($^{208}$Tl) is considered as SSE. Hence, the events are primarily between the two dashed lines. The main component of the single escape peak (SEP; 2103.51 keV) is deemed as MSE, and thus the events are mainly outside the two dashed lines. The selection of fitting energy intervals (9 gray shadow bands in the figure) is basically equidistant, avoiding the omnipotent peaks, single escape peaks, and double escape peaks of different $\gamma$-rays, because the main component of the backgrounds in the region of interest (ROI) is attributed to Compton events.
}
\label{fig:AE_E_scatter}
\end{figure}

\begin{figure}[!htbp]
\includegraphics[width=\linewidth]{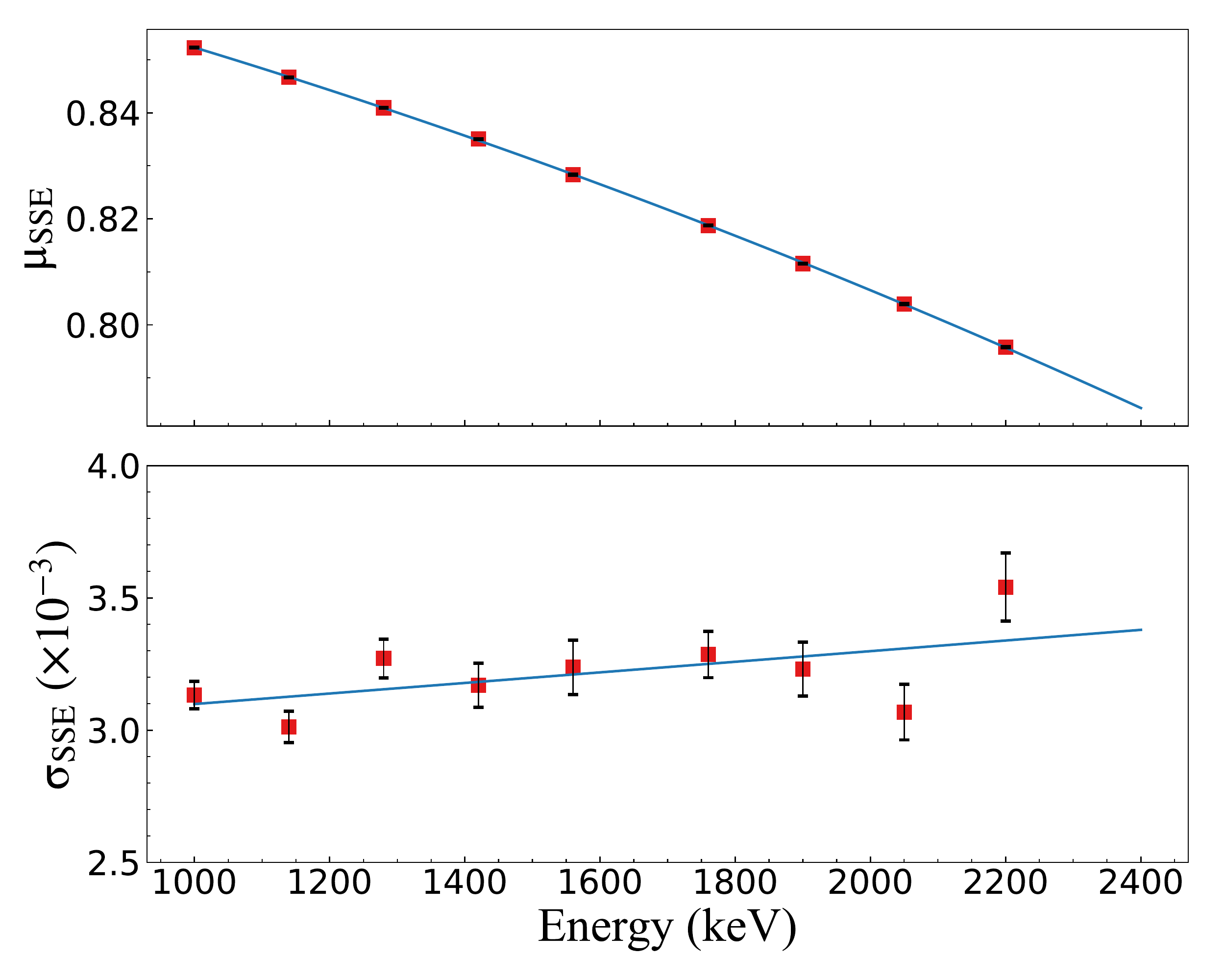}
\caption{A/E distribution parameters $\mu$ and $\sigma$ of SSE in nine energy intervals were fitted with respect to energy. Furthermore, $\mu$(E) used quadratic curve fitting, wheseas $\sigma$(E) used linear fitting, both of which were used to set the SSE/MSE discrimination threshold.}
\label{fig:sse_E_fit}
\end{figure}

\begin{figure}[!htbp]
\includegraphics[width=\linewidth]{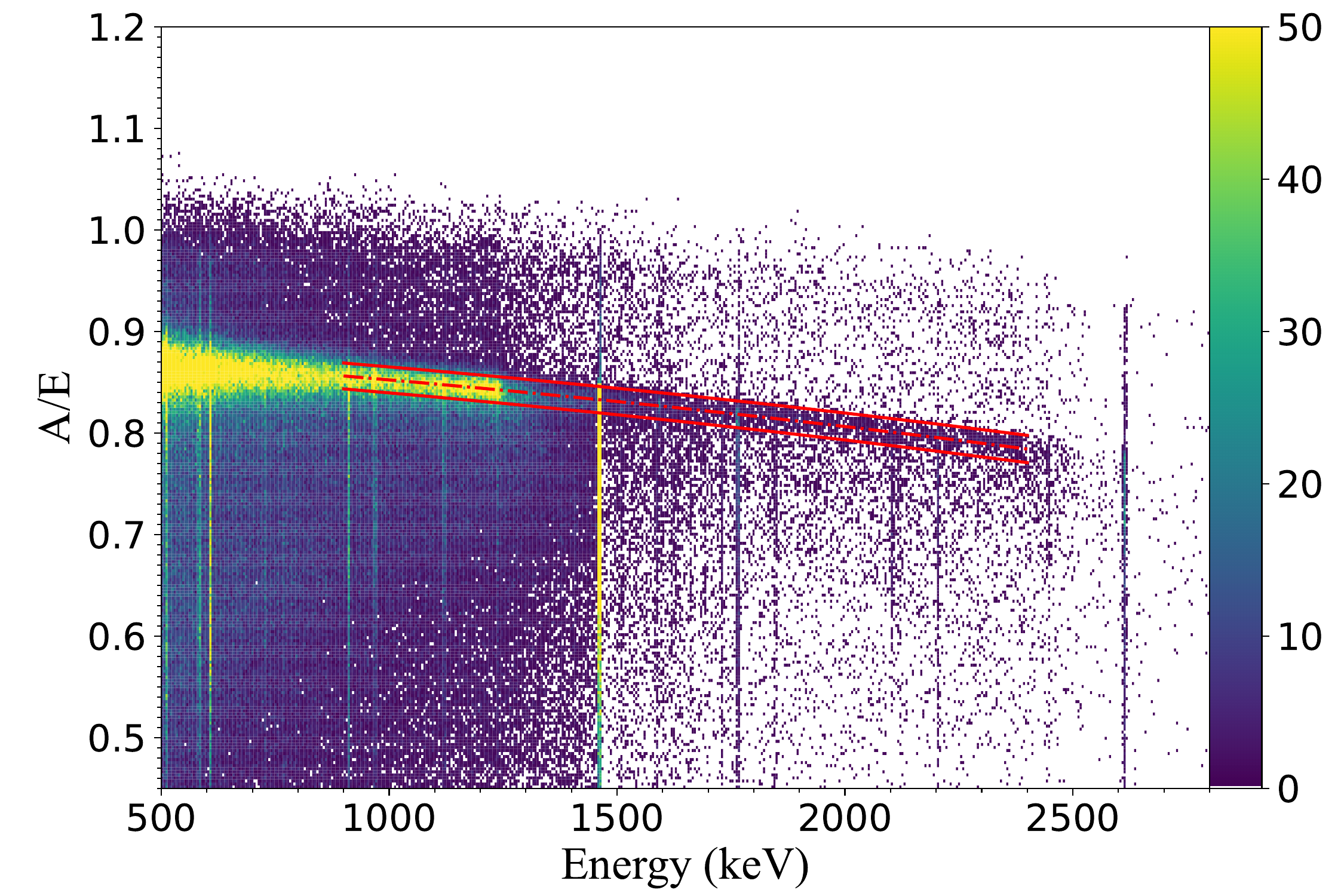}
\caption{Two-dimensional distribution diagram of A/E vs Energy of exposure data from runs IV and V. Red dashed lines denote the cut threshold.}
\label{fig:bkg_ae_e}
\end{figure}

The A/E distribution of events pertaining to a specific energy interval can be described by three components (SSE, MSE, and high A/E):
\begin{equation}
\begin{aligned}
    f(x)=&\frac{n}{\sigma  \cdot \sqrt{2\pi}} \cdot \exp{(-\frac{(x-\mu)^2}{2\sigma^2})}\\
    &+ m \cdot \frac{ \exp{(f \cdot (x-l))}+d}{\exp{(\frac{x-l}{t}}+h)} +a\cdot \sqrt{x-b},
\label{equa1}
\end{aligned}
\end{equation}
where the three terms correspond to the three components, and the distribution of the first two terms follows Ref.~\cite{GERDAPSD,GERDAPSD2022}. Furthermore, $\mu$ and $\sigma$ of events in various energy intervals were fitted using Eq.~\ref{equa1}. The energy dependence of $\mu$ and $\sigma$ is depicted in Fig.~\ref{fig:AE_E_scatter}. The A/E values across nine energy intervals (gray shadow regions in Fig.~\ref{fig:AE_E_scatter}) were fitted to obtain corresponding $\mu$ and $\sigma$. Thereafter, these $\mu$ and $\sigma$ values are fitted with the energies (midpoints of energy interval), and the energy dependence of the two parameters is depicted in Fig.~\ref{fig:sse_E_fit}. For an event with energy $E$, if its A/E value satisfies $\mu(E)-4\sigma(E)<A/E<\mu(E)+4\sigma(E)$, it will be regarded as an SSE and retained in the A/E cut. Fig.~\ref{fig:bkg_ae_e} illustrates the distribution of A/E vs Energy of exposure data from runs IV and V.

We monitored the survival fraction (SF) of the A/E cut in 1800--2200 keV energy region to evaluate the stability of the cut. A flat line was applied to fit SF via the least square method. The $\chi^2$/(degree of freedom) of the fit is 14.93/18, which can confirm the stability of the A/E parameters over long-term detector operation.

\begin{figure}[!tbp]
\includegraphics[width=\linewidth]{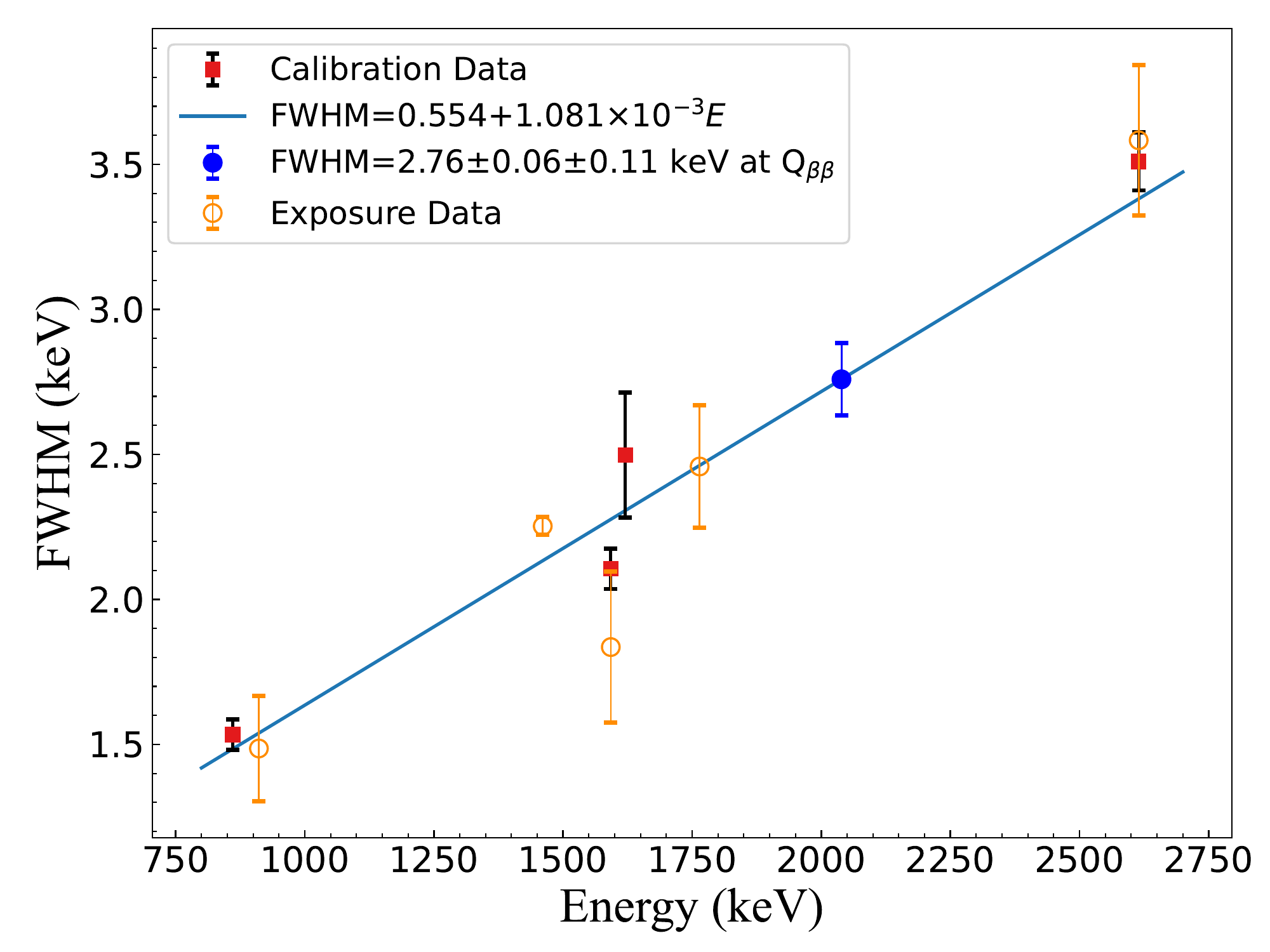}
\caption{Energy resolution of single-site events at $Q_{\beta\beta}$ derived from the $^{228}$Th calibration data. The linear fitting of the resolution was preformed with different combinations of three of the four aforementioned $\gamma$ peaks, and the maximum deviation at $Q_{\beta\beta}$ (0.11 keV) was counted as a systematic uncertainty. Uncertainty introduced by the parameters of the fitting results (0.06 keV) are considered as statistical uncertainty. The energy resolution of single-site events from exposure data is also shown for comparison.}
\label{fig:resolution}
\end{figure}

As $\bb$ events are SSEs, the energy resolution in the signal region was evaluated after A/E discrimination. Specifically, four energy peaks with sufficient statistic were selected from the $^{228}$Th calibration data, and their resolutions were fitted with respect to energy using a linear function to obtain the resolution at $Q_{\beta\beta}$ shown in Fig.~\ref{fig:resolution}. These four energy peaks correspond to 860.56 keV ($^{208}$Tl), 1592.50 keV ($^{208}$Tl), 1620.50 keV ($^{212}$Bi), and 2614.51 keV ($^{208}$Tl). The FWHMs of the four peaks are fitted with a function FWHM = $a+bE$, and the interpolation of FWHM at 2039 keV is 2.76 keV. Uncertainties of the result were derived from two aspects via the following procedure: the resolution was linearly fitted with various combinations of three of the four aforementioned $\gamma$ peaks, and the maximum deviation in the interpolated resolution at $Q_{\beta\beta}$ was accounted as a systematic uncertainty. The statistical uncertainty of FWHM at $Q_{\beta\beta}$ was calculated using the covariance matrix of the linear fit parameters ($a, b$). By combining both uncertainties, the energy resolution of SSEs at $Q_{\beta\beta}$ was derived as $2.76\pm0.13$ keV.

The CDEX-1B detector was enveloped by a NaI(Tl) anti-coincidence detector, and NaI(Tl) crystal was processed into a thick-walled well structure. During the detector operation, the low-temperature cryostat of the PPCGe detector was placed in the well~\cite{CDEX2018}. For the PPCGe detector, the enclosed solid angle was proximate to 4$\pi$, resulting in a relatively high anti-coincidence efficiency~\cite{CDEX2018}. As the background events in region of interest (ROI) are primarily composed of Compton scattering events, there exists a high probability of depositing energy outside the detector. The anti-coincidence detector exhibited a strong coincident ability for background events in the high-energy region and reduced the background level in the ROI by nearly an order of magnitude.

\begin{figure}[!htbp]
\includegraphics[width=\linewidth]{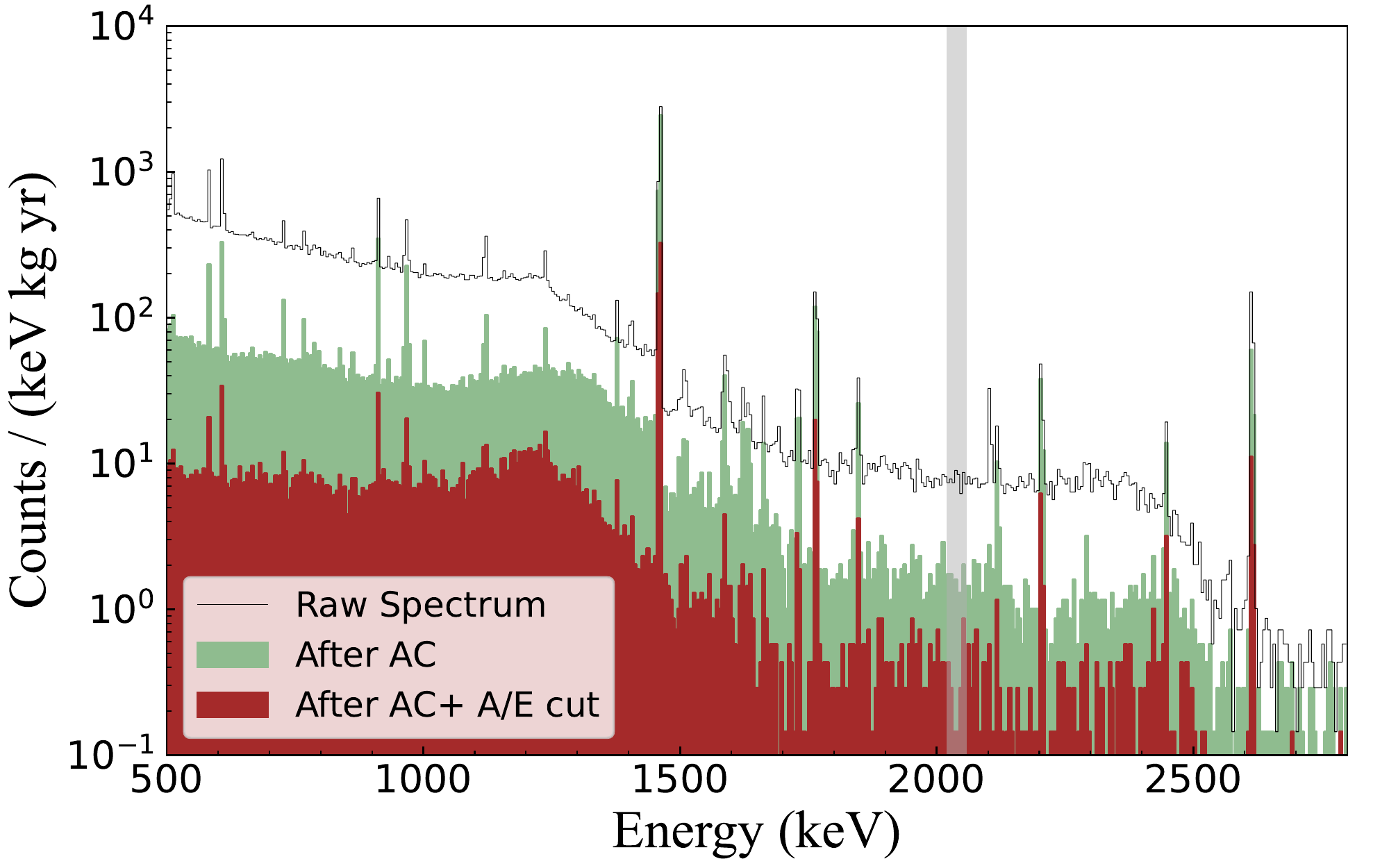}
\caption{Energy spectra of the CDEX-1B detector (500--2800 keV, 504.3 kg$\cdot$day); the shadow band represents the signal region of $^{76}$Ge $Q_{\beta\beta}$. The region of interest (ROI) background level before the anti-coincidence and MSE/SSE cut was approximately 7.7 counts/(keV$\cdot$kg$\cdot$yr). After the anti-coincidence (AC) cut, the ROI background level decreased by nearly one order of magnitude. After the AC and MSE/SSE cuts, the magnitude of the ROI background level was 0.33 counts/(keV$\cdot$kg$\cdot$yr). The combination of these two cuts significantly suppressed on the background.}
\label{fig:C1B_bkg_spec}
\end{figure}

As depicted in Fig.~\ref{fig:C1B_bkg_spec}, the background level in ROI is suppressed from 7.7 to 0.33 counts/(keV$\cdot$kg$\cdot$yr). This corresponds to a 23-fold reduction, achieved through anti-coincidence (AC) and A/E cuts.

Previous calibration experiments demonstrated that the dead-layer thickness of the CDEX-1B detector was $0.88\pm 0.12$ mm~\cite{MJL2017}. The energy of a $\bb$ event occurring at the edge of the detector (near the dead layer) may not be fully collected, thereby reducing the detection efficiency of $\bb$ events. To calculate the efficiency, we used Geant4 for simulating the $\bb$ events in the CDEX-1B germanium crystal and sampled the emission directions as well as the energies of the two electrons according to the following Eq.~\ref{eq:sample}~\cite{0vbbSample}:
\begin{align}
F(T_1,T_2,\cos{\theta})=&(T_1+1)^2(T_2+1)^2\times \nonumber\\
&\delta (T_0-T_1-T_2)(1-\beta_1\beta_2\cos{\theta}),
\label{eq:sample}
\\
\beta_i=&\frac{\sqrt{T_i(T_i+2)}}{T_i+1},
\end{align}
where $T_0$ denotes $Q_{\beta\beta}$, $T_i$ indicates the electron kinetic energy, and $\theta$ represents the angle between the emission directions of two electrons.
The proportion of $\bb$ event with complete energy deposition (at the energy range of $Q_{\beta\beta} \pm{3\sigma}$) can be calculated as $(84.8\pm{0.8})\%$.

The $\bb$ signal survival efficiency of the MSE/SSE discrimination is a key parameter. Theoretically, $\bb$ events are uniformly distributed in the detector crystal. The $^{228}$Th radiation source used in the calibration experiment is a point-like source external to the Ge crystal, resulting in a non-uniform distribution of the induced double escape peak (DEP) events. Particularly, given that we put the source close to the electrode surface, where the gradient of weight potential field is large, the fraction of high A/E DEP events is very high, which leads to a low survival efficiency. Therefore, the survival efficiency of $\bb$ events cannot be directly derived from the calibration data. Consequently, we applied a simulation method~\cite{GERDAsimu}, the accuracy of which was verified by the calibration data.

The simulation included three steps: 
\begin{itemize}
  \item Simulation of energies and positions using Geant4 software~\cite{GEANT4}.
  \item Simulation of the carrier drift processes and the pulses generated through the pulse shape simulation package in SAGE~\cite{SAGE} based on the Shockley-Ramo theory~\cite{HE2001250}. These original pulses were convolved with electronics response function, which was obtained by inputting a $\delta$-like current pulse with a very fast rising edge into the amplifier chain. The noise in the experimental pulses was superimposed onto the simulated pulse.
  \item Extraction of energy and A/E parameters in simulated pulses via the aforementioned pulse-processing procedures.
\end{itemize}

\begin{figure*}[!htbp]
\includegraphics[width=\linewidth]{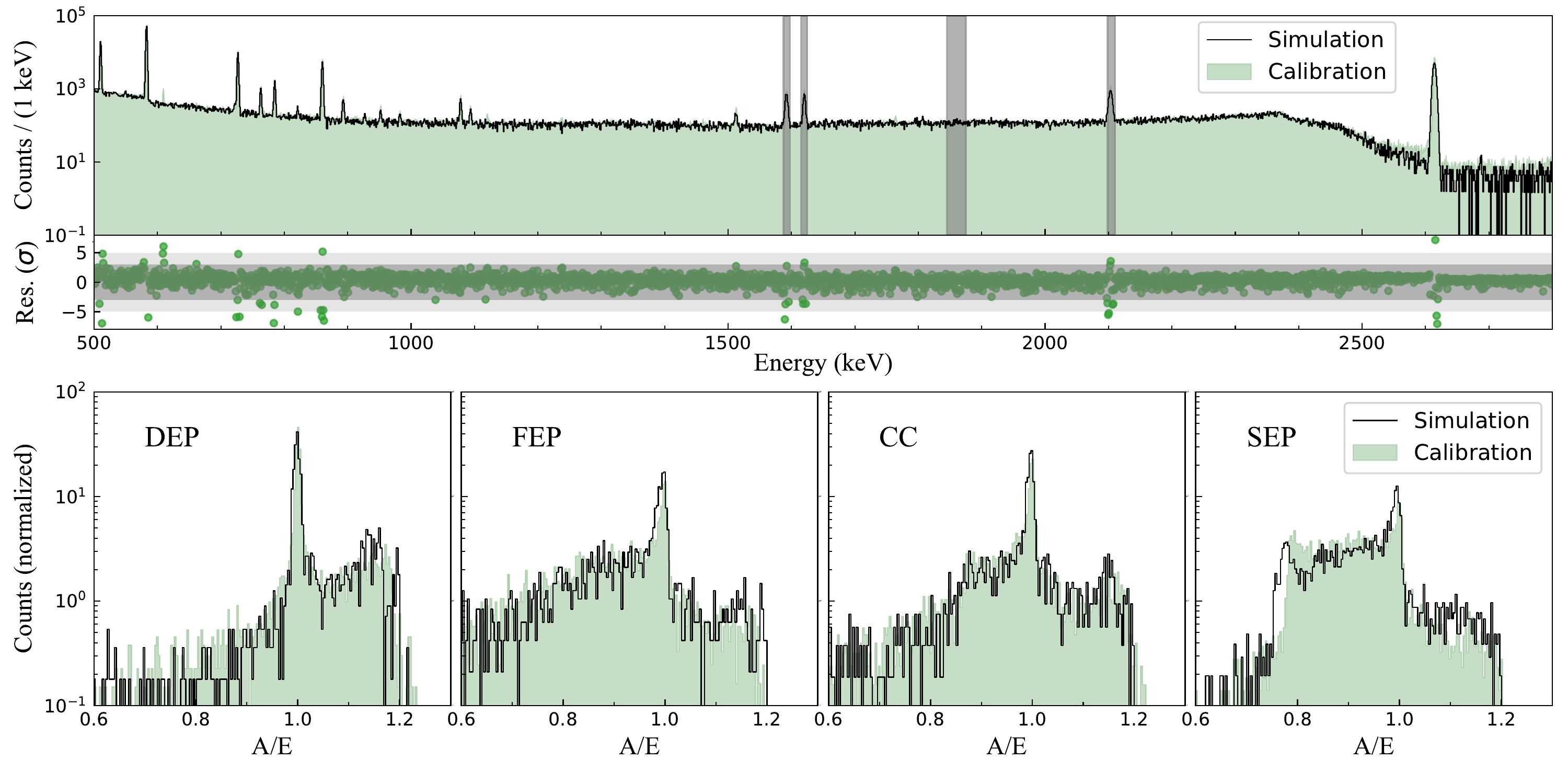}
\caption{(Top) Calibration and normalized simulated spectra of $^{228}$Th. Shadow bands include four distinct types of energy regions: double escape peak (DEP, 1592.5 keV), full energy peak (FEP, 1620.5 keV), Compton continuum (CC, 1860.0 keV), and single escape peak (SEP, 2103.5 keV). Calibration and simulated A/E distributions were compared in these energy regions. (Bottom) Simulated A/E distributions (after normalization, the center value of SSE distribution is 1) of $^{228}$Th are consistent with the corresponding calibration data. The difference between the DEP's cut efficiencies of the simulated data and calibration data was regarded as a systematic uncertainty of the A/E cut survival efficiency of $\bb$ events.}
\label{fig:MC_cmp}
\end{figure*}

As illustrated in Fig.~\ref{fig:MC_cmp}, the A/E distributions of the simulated events in different energy intervals are in good agreement with the calibration data. The survival rates of DEP, FEP, and SEP events in Fig.~\ref{fig:MC_cmp} under the $3\sigma$, $4\sigma$, and $5\sigma$ cut thresholds are compared with the experimental results presented in Table \ref{tab:threashold}. The difference of DEP ($\pm$3.8\%) was considered as a systematic uncertainty. The same method was used to simulate the A/E distribution of the $\bb$ events, which were uniformly distributed in the Ge crystal. Finally, the proportion of $\bb$ events passing the A/E cut (A/E cut survival efficiency of $\bb$ events) was calculated as $(78.6\pm 3.8)\%$. 

\begin{table}
  \centering
  \caption{Comparison of A/E cut efficiency for DEP, FEP, and SEP events between calibration and pulse shape simulation.}
\begin{tabular}{cccccc}
\hline
\hline Event&Cut threshold & Calibration & Simulation & Diff. & Uncertainty \\
\hline 
&$3 \sigma$ & $46.38 \%$ & $51.00 \%$ & $4.62 \%$ & \\
DEP&$4 \sigma$ & $49.17 \%$ & $52.56 \%$ & $3.39 \%$ & $3.8 \%$ \\
&$5 \sigma$ & $51.15 \%$ & $54.36 \%$ & $3.22 \%$ & \\
\hline 
&$3 \sigma$ & $17.72 \%$ & $24.75 \%$ & $7.03 \%$ & \\
FEP&$4 \sigma$ & $20.22 \%$ & $29.06 \%$ & $8.84 \%$ & $8.5 \%$ \\
&$5 \sigma$ & $22.21 \%$ & $31.73 \%$ & $9.50 \%$ & \\
\hline 
&$3 \sigma$ & $13.54 \%$ & $17.83 \%$ & $4.29 \%$ & \\
SEP&$4 \sigma$ & $15.88 \%$ & $21.10 \%$ & $5.22 \%$ & $5.3 \%$ \\
&$5 \sigma$ & $17.75 \%$ & $24.17 \%$ & $6.42 \%$ & \\
\hline
\hline
\end{tabular}
\label{tab:threashold}
\end{table}

Moreover, the CDEX-1B detector was equipped with a pulsed reset preamplifier. The preamplifier resets after the voltage reaches the threshold owing to the continuous and stable leakage current. The charge collection of each event decreases the voltage, and the decreased amplitude is directly proportional to the collected charge (corresponding to the event energy)~\cite{CDEX2017}. Events with higher energy are more likely to decrease the voltage to reach the threshold, and thereby, resetting the amplifier.  These types of events are measured with degraded energy and have been excluded from the analysis. Under the low counting rate condition ($\sim$1 count/period), the detection efficiency can be expressed as follows:
\begin{align}
\eta(E)= 1- \frac{E}{E_s-E_s\cdot \frac{T_{veto}}{T_0}} 
\approx 1-\frac{E}{E_s}(\frac{T_{veto}}{T_0}\ll 1),
\label{equa4}
\end{align}
where $E$ denotes the energy of the event and $E_s$ denotes the saturation energy of the preamplifier. Any event with energy $E_s$ will cause a voltage decrease equal to the voltage decrease during a entire period of preamplifier. In this study, $E_s$ was evaluated as 4.7 MeV. A period of a few milliseconds exists after a reset during which the baseline is distorted. Specifically, $T_{veto}$=10 ms is a preset time window to veto the signals in this period, and $T_0\approx$ 620 ms is the reset period if there is no signal in the period. Upon fitting the data during the entire operation time, the efficiency at $\bb$ $Q_{\beta\beta}$ energy was (56.6 $\pm{}$ 0.1)\%.

\section{Background model}
Understanding the spectral profile of background events in the ROI is necessary. We used Geant4 to simulate the energy spectra generated by the background sources. Detector components, including the copper shell surrounding the crystal, lead foil, PTFE tube, electrode components, PMT, NaI(Tl) crystal, and copper shell outside the crystal, are considered in the simulation. The energy spectra of $^{238}$U chain, $^{232}$Th chain, and $^{40}$K in these components were simulated. Furthermore, for the lead foil, we simulated the energy spectrum of $^{210}$Pb. For the gap between NaI(Tl) and HPGe detectors, we simulated the energy spectrum of $^{222}$Rn and its decay daughters, assuming a secular equilibrium in the decay chain. Accordingly, we considered cosmogenic isotopes in the Ge crystal including $^{60}$Co, $^{57}$Co, and $^{68}$Ga. Notably, after years of underground operations, the cosmogenic isotopes in the CDEX-1B detector have been reduced to a negligibly low level. 

\begin{figure}[!htbp]
\includegraphics[width=\linewidth]{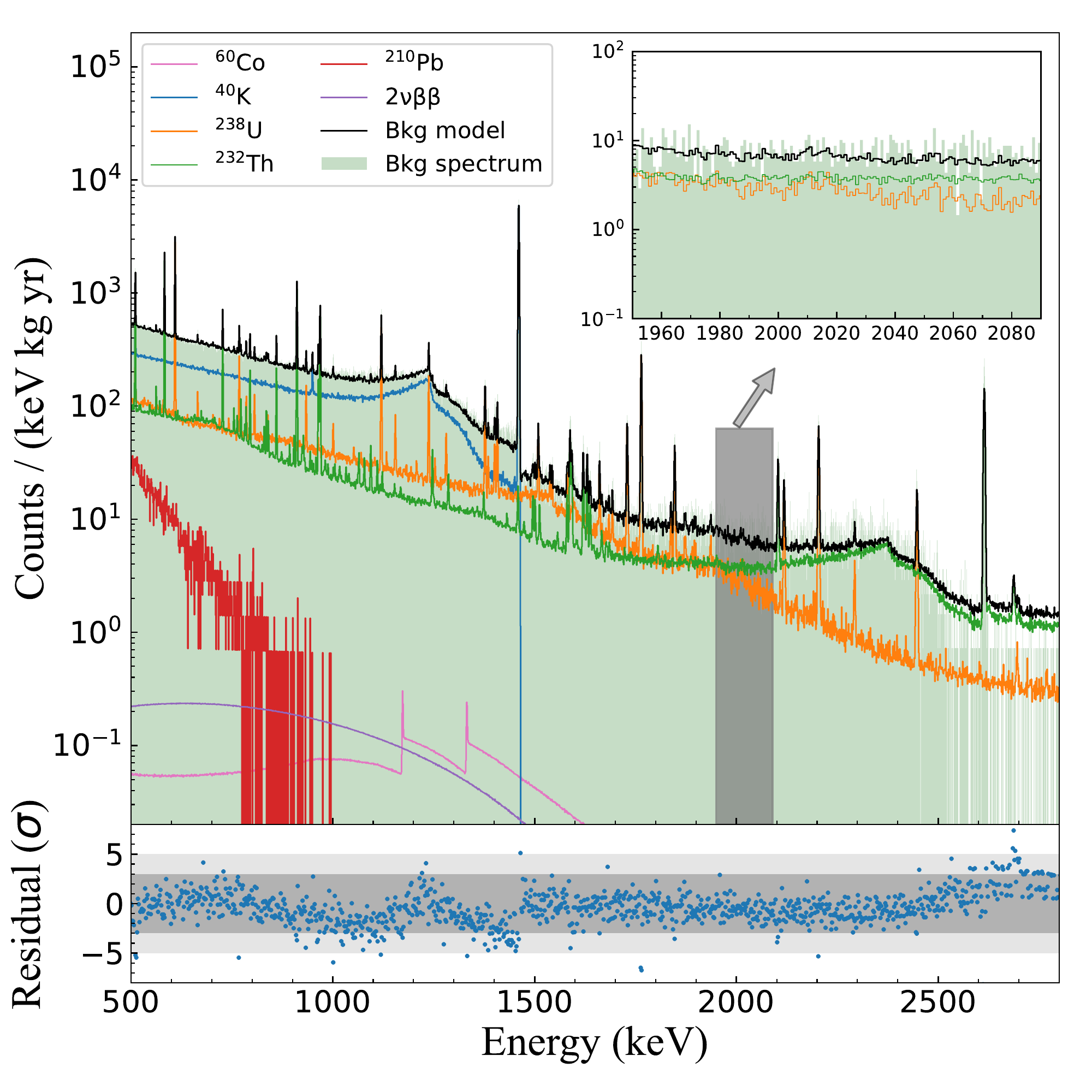}
\caption{Energy spectrum component analysis of CDEX-1B; the green area shows the raw measured energy spectrum without events-selection cuts; the thin black line indicates the fitting result of the background model. The background in the ROI energy region mainly originates from the $^{232}$Th decay chain and the $^{238}$U decay chain of the external components of the CDEX-1B germanium crystal and cosmogenic nuclide $^{60}$Co inside the germanium crystal. The magnified view of the shaded area shows a flat profile of the background spectrum in the ROI.}
\label{fig:C1B_bkg_model}
\end{figure}

After simulating all potential background sources, we selected the 500--2800 keV energy region and obtained a background model by fitting the experimental data with the simulated spectra using a chi-square method. Owing to the high similarity among the energy spectra of the same nuclides, the energy spectra of the same nuclides were merged across various components in the background model for fitting, and a simplified background model was obtained for the energy spectra of various nuclides. In the ROI, the main contribution of the background originates from the $^{232}$Th chain, followed by the $^{238}$U chain. Figure~\ref{fig:C1B_bkg_model} shows that the background model is in good agreement with the measured energy spectrum, and the normalized residual value in each bin is basically within 3$\sigma$. Furthermore, the background model exhibited a flat profile of the background in the ROI energy region, which will be considered in the subsequent physical analysis.

\section{Results and discussion}
\begin{figure}[!tbp]
\includegraphics[width=\linewidth]{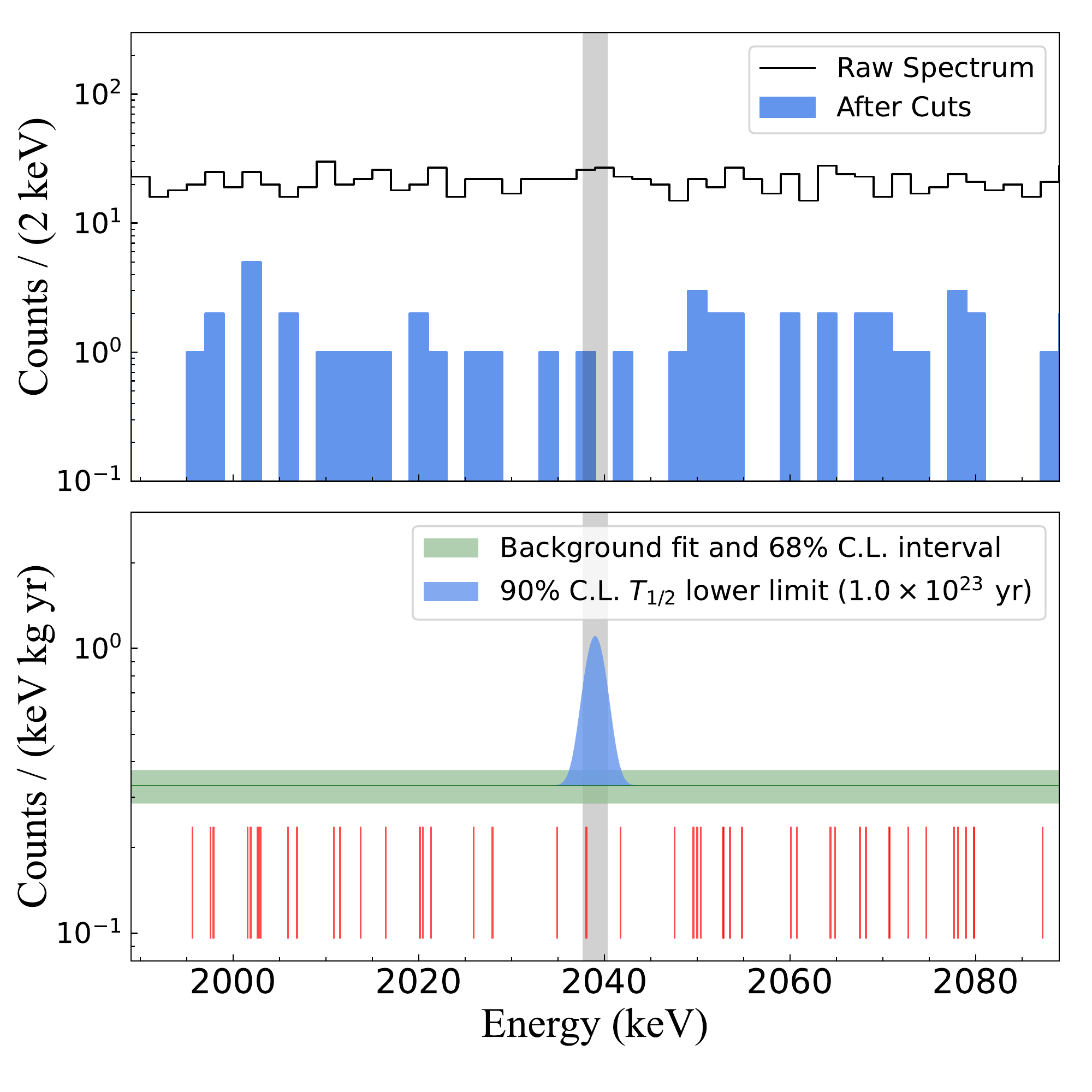}
\caption{Physical results of the $^{76}$Ge $\bb$ analysis from the CDEX-1B experiment. (Top) Energy spectrum in the energy interval of [1989 keV, 2089 keV] with a bin width of 2 keV, where no peak from the background is observed, and a flat background is assumed. Shadow interval denotes the FWHM energy region of [2037.6 keV, 2040.4 keV]. (Bottom) Best-fit result of the background level and its 68\% confidence interval (Poisson distribution). The blue Gaussian peak represents the signal strength of the 90\% confidence level, corresponding to the lower limit of $\bb$ half-life, $T_{1/2}^{0\nu}\ > \ {1.0}\times 10^{23}\ \rm yr\ (90\% \  C.L.)$. Each thin red line indicates a single event in the energy region after analysis cuts.}
\label{fig:C1B_0vbb_result}
\end{figure}

By using a total of 504.3 kg$\cdot$day effective exposure data, the energy interval of [1989 keV, 2089 keV] was selected to analyze the half-life of $^{76}$Ge $\bb$ (Fig.~\ref{fig:C1B_0vbb_result}). The relationship between the expected signal count, $\mu_s$, and half-life $\Thf{}$ can be determined as follows:
\begin{equation}
\begin{aligned}
\mu_{s} = \frac{1}{T_{1/2}^{0\nu}} \cdot \frac{N_A \cdot \ln{2} }{m}\cdot \epsilon \cdot  n \cdot \eta,
\label{equa5}
\end{aligned}
\end{equation}
where $N_A$ denotes the Avogadro's constant, $m$ represents the molar mass of natural Ge, $\epsilon$ indicates the exposure (504.3 kg$\cdot$day), $n$ denotes the $^{76}$Ge abundance (7.83\%), and $\eta$ represents the total efficiency of detecting $\bb$ events, which is a product of the inhibit efficiency ($(56.6\pm{0.1})\% $), anti-coincidence efficiency ($99.8\%$), MSE/SSE cut efficiency ($(78.6\pm{3.8})\% $), and $\bb$ event energy complete deposition ratio ($(84.8\pm{0.8})\% $).

Based on the assumption of the flat background in the ROI, the expected background count of the analyzed region can be determined as follows:
\begin{equation}
\begin{aligned}
\mu_{b}=B \cdot \Delta E \cdot \epsilon,
\label{equa6}
\end{aligned}
\end{equation}
where $\Delta E = 100$ keV, and $B$ represents the background index (counts/(keV$\cdot$kg$\cdot$yr)).
 
After comprehensive analysis cuts, only 46 events remained in the analyzed energy region. Accordingly, the extended likelihood function was utilized based on the unbinned method for analysis~\cite{GERDAphy,PDG2022}. The likelihood function is formulated as
\begin{equation}
\begin{aligned}
L(S,B)& = \frac{(\mu_s+\mu_b)^N e^{-(\mu_s+\mu_b)}}{N!}\times \\ 
&\prod_{i=1}^{N}
\frac{1}{\mu_s+\mu_b}\times(\frac{\mu_b}{\Delta E}+\frac{\mu_s}{\sqrt{2\pi}\sigma} e^{-\frac{(E_i-Q_{\beta\beta}-\delta_{E})^2}{2\sigma^2}}),
\label{equa7}
\end{aligned}
\end{equation}
where $E_i$ denotes the energy of the $i$-th event, {$\delta_{E}$ is a possible systematic energy offset at $Q_{\beta\beta}$},  $\mu_s$ is the expected observed signal event number, $\mu_b$ is the expected background event number, $N$ is the observed event number, and $\sigma=\rm{FWHM/(2\sqrt{2ln2})}$ is the energy resolution of SSEs in the ROI. As predicted by the background model, a flat background was assumed in the analyzed region. For the signal, a Gaussian distribution centered at the $Q_{\beta\beta}$ with a width corresponding to the energy resolution was considered.

The constraints on the signal number were derived using frequentist analysis and a two-side test statistic~\cite{GERDAphy,Cowan2011}:
\begin{equation}
\begin{aligned}
t_S=-2\ln{\frac{L(S,\hat{\hat{B}})}{L(\hat{S},\hat{B})}},
\label{equa8}
\end{aligned}
\end{equation}
where $S$ denotes the strength of a possible $\bb$ signal: $S= 1/T_{1/2}^{0\nu}$, corresponding to the expected event number $\mu_s$, and $\hat{\hat{B}}$ denotes the background index obtained by maximizing $L$ when $S$ is fixed. Parameters $\hat{S}$ and $\hat{B}$ correspond to the global maximum $L$. The unbinned profile likelihood analysis yields a best-fit background of $\hat{B}=0.33$ counts/(keV$\cdot$kg$\cdot$yr) and signal strength of $\hat{S}=0$ yr$^{-1}$ indicating no signal. Accordingly, we selected a discrete set of $S\in \{S_j\}$ for simulation. For each $S_j$, multiple Monte Carlo simulations were performed in combination with other experimental parameters. Furthermore, $t_{S_j}$ was calculated using the obtained energy spectrum, and its probability distribution $f(t_s|S_j)$ was obtained through repeated simulations~\cite{GERDAphy}. The $t_{obj}$ value was calculated using the experimental energy spectrum. The $P$ value of $S_j$ and observed data were calculated as follows:
\begin{equation}
\begin{aligned}
P_{S_j}=\int_{t_{obs}}^{\infty}f(t_s|S_j){\rm d} (t_{S_j}).
\label{equa9}
\end{aligned}
\end{equation}

The lower limit for $^{76}$Ge $\bb$ half-life corresponding to $P=0.1$ is set to:
\begin{equation}
\begin{aligned}
T_{1/2}^{0\nu}\ > \ {1.0}\times 10^{23}\ \rm yr\ (90\% \  C.L.).
\label{equa10}
\end{aligned}
\end{equation}

The expectation of the experimental sensitivity was evaluated based on the distribution of $P_{S_j}$ which was derived from Monte Carlo generated data sets assuming no $\bb$ signal ($S=0$)~\cite{GERDAphy}. The median sensitivity was calculated as $5.6\times10^{22}$ yr$\ (\mathrm{ 90\% \ C.L.})$. The probability of obtaining a limit stronger than the actual limit is 31\%.

 The upper limit of $\bb$ signal strength corresponding to the above half-life ($1.0\times10^{23}$ yr) is 2.28 events. The energy resolution (2.76 $\pm$ 0.13 keV) is considered by folding it into the profile likelihood function through additional nuisance parameters constrained by Gaussian probability distribution. Uncertainty of the overall efficiency (37.6 $\pm$ 1.9\%) is considered by propagating it through Eq.~\ref{equa5}. The overall effect of the uncertainty on the half-life limit is 4.9\%. We used Eq.~\ref{eq:mbb} to convert the limit on $T_{1/2}^{0\nu}$ to the limits on effective Majorana neutrino mass, $\langle m_{\beta\beta}\rangle$:
\begin{equation}
    \langle m_{\beta\beta}\rangle=\frac{m_e}{|g_A^2\cdot M^{0\nu}|\cdot \sqrt{T_{1/2}^{0\nu}\cdot G^{0\nu}}},
    \label{eq:mbb}
\end{equation}
where the phase space factor $G^{0\nu}$ for $^{76}$Ge is ${2.36}\times 10^{-15}\ \mathrm{yr}^{-1}$~\cite{Kotila2012}, and the coupling constant $g_A = 1.27$~\cite{GERDA2020}. Given the values of nuclear matrix element $M^{0\nu}$, calculated using different theoretical models, differ considerably, we regarded it as an interval, with a range of 2.66--6.34, selected from Refs.~\cite{Horoi,Barea,Juhani,Fedor,Nuria,Yao,Tomas,Coraggio2020,PhysRevD.102.095016}.
The upper limit on the effective Majorana neutrino mass obtained in this study corresponds to $\langle m_{\beta\beta}\rangle <$ 3.2--7.5$\  \mathrm{eV}\ (\mathrm{ 90\% \ C.L.})$.

\section{Summary}
In this study, we explored the use of a PPCGe in the CDEX-1B experiment for $^{76}$Ge $\bb$ decay detection. The PPCGe detector is originally designed for dedicated low-energy dark matter searches. However, we upgraded its preamplifier and applied the PSD and anti-veto background suppression technologies to suit the search for $\bb$ signal. 

The CDEX-1B PPCGe detector realized an energy resolution of 2.8 keV in the $\bb$ signal region, a 23-fold background reduction using PSD and anti-veto, a smooth operation of over 500 days, and provided a $^{76}$Ge $\bb$ decay half-life limit of $1.0\times10^{23}\ \mathrm{yr}$ (90\% C.L.). These performances exceeded the previous CDEX $\bb$ studies~\cite{CDEX2017,DWH2022} (as in Table~\ref{tab:0vbbEeff}) and demonstrated that the PPCGe detectors designed for dark matter search could be used in future large-scale $\bb$ experiments.

\begin{figure}[!bp]
\includegraphics[width=\linewidth]{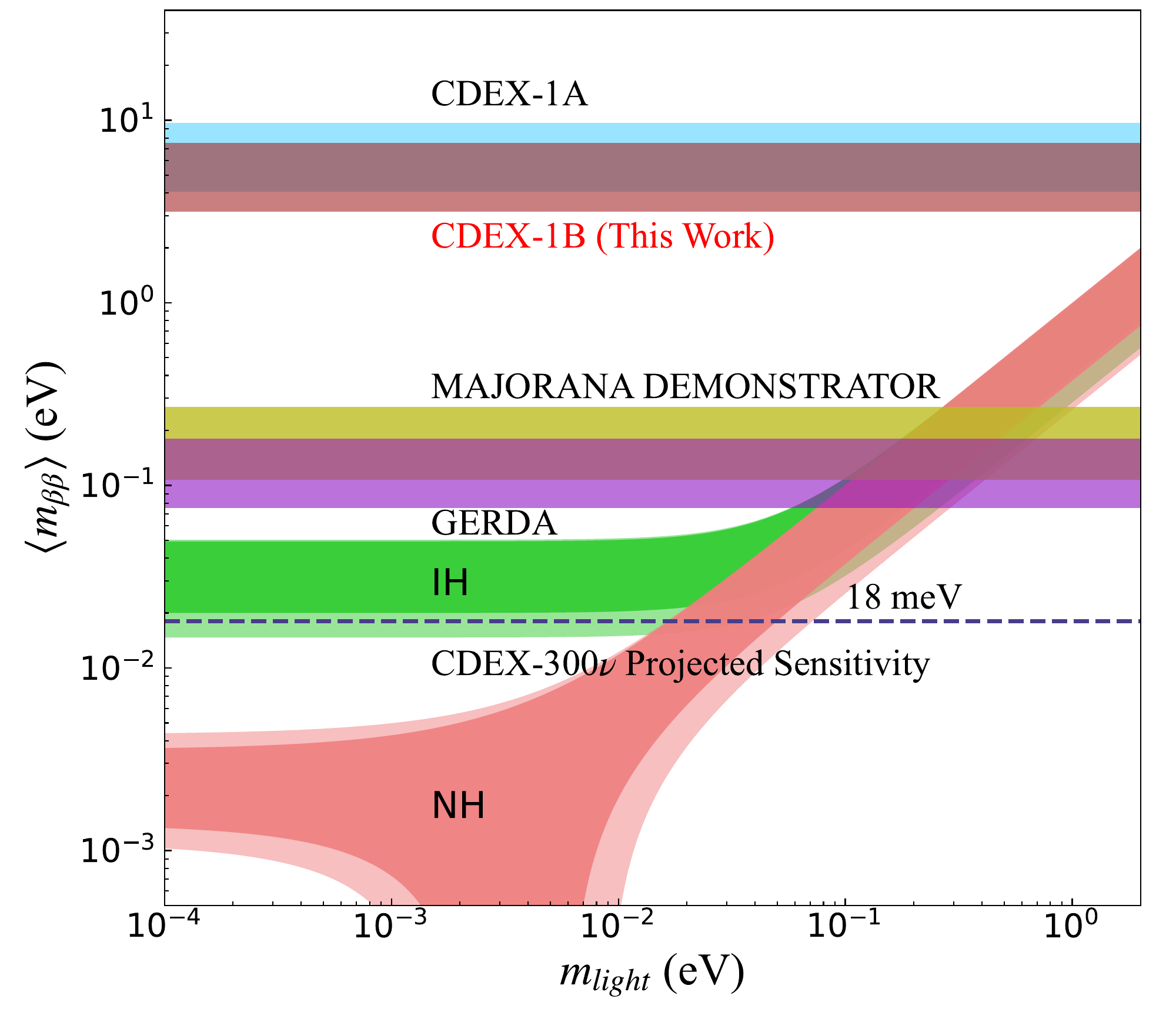}
\caption{CDEX-1B result of the effective Majorana neutrino mass. The green band and magenta band indicate the parameter space  allowed by the inverted hierarchy (IH) and normal hierarchy (NH) as a function of the lightest neutrino mass $m_{light}$, respectively~\cite{PDG2022}. The brown and blue bands denote the result from CDEX-1B (this work) and CDEX-1A experiments~\cite{CDEX2017}, respectively. The purple and yellow bands represent the final result of the GERDA and M\textsc{ajorana} D\textsc{emonstrator} experiments, respectively~\cite{GERDA2020,MJD2023}. Currently, these two experiments hold leading positions in the $^{76}$Ge-based $\bb$ experiments. Furthermore, the projected sensitivity of CDEX-300$\nu$ is imposed. All the effective Majorana neutrino mass is calculated using the latest values of nuclear matrix element, 2.66--6.34~\cite{RevModPhys.95.025002}.}
\label{fig:0vbb_Mbb_C1Bwork}
\end{figure}

\begin{table*}
  \centering
  \caption{Comparison of the experimental parameters and results~\cite{CDEX2017,DWH2022}}
  {
\begin{tabular}{lccc}
\toprule
\hline\hline
 & CDEX-1B (This work) & CDEX-1A & BEGe experiment  \\
 \hline
\midrule
Exposure (kg$\cdot$day)  & 504.3 & 304 & 186.4\\
Energy resolution (keV, FWHM@$Q_{\beta\beta}$) &$2.76\pm0.13  $&$4.3 \pm0.2  $& $2.85 \pm0.48$\\
Efficiency       & 37.7\%     & 68.4\% & 73.2\% \\
Background index (counts/(keV$\cdot$kg$\cdot$yr)) &$0.33\pm0.05  $  & 4.38 &$2.35\pm 0.11$\\
$T_{1/2}^{0\nu} $\ limit (90\% \ C.L.)  &$1.0\times 10^{23}~\mathrm{yr}$  &   $6.4\times 10^{22}~\mathrm{yr}$ &$5.6\times 10^{22}~\mathrm{yr}$              \\
\hline\hline
\bottomrule
\end{tabular}}
\label{tab:0vbbEeff}
\end{table*}

We noted that the current background is approximately at the level of dedicated $\bb$ experiments using HPGe detector in the late 80s of last century~\cite{rev2019}, still far from the state-of-art~\cite{GERDA2020} and requirement of future experiments. The background model shows radionuclides in $^{238}$U and $^{232}$Th decay chains, and they are the primary sources of background in the $\bb$ signal region. These sources are mainly from the surrounding materials. Therefore, a stringent material screening is crucial for future low-background experiments.

The next-generation CDEX $\bb$ project, CDEX-300$\nu$, targets at searching $\bb$ signal at the inverted hierarchy (IH) region with a sensitivity reaching the bottom of the IH region (Fig.~\ref{fig:0vbb_Mbb_C1Bwork}). The CDEX-300$\nu$ will operate approximately 225 kg of enriched germanium detectors ($^{76}$Ge abundance $>$ 86\%) coupled with a liquid argon (LAr) system serving as the cooling medium and veto detector at CJPL~\cite{DWH2022}. All materials used in detector structure and experiment construction will be screened to control their backgrounds. Furthermore, in light of the enhanced background suppression power by anti-veto and PSD in this study and the GERDA experiment~\cite{GERDA2020}, we expect to have strong background suppression power in CDEX-300$\nu$ with LAr as the veto detector. Furthermore, existing PPCGe detectors used in dark matter searches can also be added into the detector array. Those PPCGe will serve as the detectors for $\bb$ while maintaining their ability to search dark matter and other new physics in the low energy region.

The CDEX-300$\nu$ experiment is planned to realize a background level of $1\times 10^{-4}$ counts/(keV$\cdot$kg$\cdot$yr), and it is expected to attain a half-life sensitivity of $3.3\times 10^{27}$ yr for $^{76}$Ge $\bb$ in 10 years of effective runtime, corresponding to the effective Majorana neutrino mass region of $\langle m_{\beta\beta}\rangle <$ 18--43$\ \mathrm{meV}\ (\mathrm{ 90\% \ C.L.})$ (Fig.~\ref{fig:0vbb_Mbb_C1Bwork}).

\section*{Acknowledgements}
 We are grateful to China Jinping Underground Laboratory (CJPL) and its staff for hosting and supporting the CDEX project. CJPL is jointly operated by Tsinghua University and Yalong River Hydropower Development Company.

B. T. Zhang and J. Z. Wang contributed equally to this study.

\bibliography{C1B_0vbb.bib}

\end{document}